# A Method to Compute the Sparse Graphs for Traveling Salesman Problem Based on Frequency Quadrilaterals

Yong Wang[1] and Jeffrey Remmel[2]

[1] North China Electric Power University, Beijing 102206, China
[2] University of California, San Diego, California, 92093, USA
wangyyong100@163.com

**Abstract.** In this paper, an iterative algorithm is designed to compute the sparse graphs for traveling salesman problem (*TSP*) according to the frequency quadrilaterals so that the computation time of the algorithms for *TSP* will be lowered. At each computation cycle, the algorithm first computes the average frequency $\bar{f}(e)$ of an edge $e$ with $N$ frequency quadrilaterals containing $e$ in the input graph $G(V,E)$. Then the $1/3|E|$ edges with low frequency are eliminated to generate the output graph with a smaller number of edges. The algorithm can be iterated several times and the original optimal Hamiltonian cycle is preserved with a high probability. The experiments demonstrate the algorithm computes the sparse graphs with the $O(n\log_2 n)$ edges containing the original optimal Hamiltonian cycle for most of the *TSP* instances in the *TSPLIB*. The computation time of the iterative algorithm is $O(Nn^2)$.

## 1 Introduction

Traveling Salesman Problem (*TSP*) is a well-known *NP*-hard problem in combinatorial optimization. Given the complete graph $K_n$ on the $n$ vertices $\{1, 2, \cdots, n\}$, there is a distance function $d(u, v) > 0$ between any pairwise vertices $u, v \in \{1, 2, \cdots, n\}$. For the symmetric *TSP*, we have $d(u, v) = d(v, u)$. The objective of *TSP* is to find such a permutation $\sigma = (\sigma_1, \sigma_2, \cdots, \sigma_n)$ of the $n$ vertices $\sigma_k \in \{1, 2, \cdots, n\}(1 \leq k \leq n)$ where the total distance $d(\sigma) = d(\sigma_1, \sigma_n) + \sum_{k=1}^{n-1} d(\sigma_k, \sigma_{k+1})$ is the minimum. Namely, the cycle $\sigma = (\sigma_1, \sigma_2, \cdots, \sigma_n)$ with the minimum distance $d(\sigma)$ is the optimal Hamiltonian cycle (*OHC*) and the other cycles $\sigma$s are called the Hamiltonian cycles (*HC*). *TSP* has been proven to be *NP*-complete [1] and there are no exact polynomial-time algorithms unless $P = NP$.

*TSP* is the ultimate model of many complex discrete optimization problems, such as the network optimization, VLSI, and machine scheduling, etc. The methods for *TSP* are usually referred to resolve these complicate problems. Thus, *TSP* is extensively studied in combinatorics, operation research and computer science, etc. The algorithms for *TSP* [2] have become one of the prosperous branches in the research. The research illustrated that the exact algorithms generally need $O(a^n)$ time to resolve *TSP* where $a > 1$. For example, the time complexity of dynamic programming for *TSP* is $O(n^2 2^n)$ owing to Held and Karp [3], and independently Bellman [4]. The state-of-art branch and bound [5] or cutting plane methods [6, 7] are feasible for *TSP* with thousands of vertices. Due to the exponential number of constraints, the exact methods for *TSP* often need long-time computation to resolve the big scale of *TSP* instances.

Since the computation time of the exact algorithms is hard to reduce, some researchers turn to the approximation algorithms or heuristics for *TSP*. The approximation algorithms mainly depend on the good properties of some special computation models, such as special graphs or trees, to reduce the computation time. The minimum spanning tree-based algorithm [8] and Christofide's algorithm [9] spend the $O(n^2)$ and $O(n^3)$ time to produce the 2-approximation and 1.5-approximation for metric *TSP*, respectively. The computation time of the approximation algorithms usually have close relationships with the approximation ratio. The nearer the approximation approaches the optimal solution, the longer the computation time of the approximation algorithms will require [2]. The experiments illustrated that the Lin−Kernighan heuristics (*LKH*) was competitive to generate the approximation [10] within 5% of the optimal solution in nearly $O(n^{2.2})$ time. One sees the heuristic algorithms are efficient to compute the approximations. On the other hand, they cannot guarantee to find the optimal solution, especially for the large scale of *TSP*.

Besides the above methods for *TSP* on the $K_n$, some researchers pursue the methods for *TSP* on the sparse graphs. The sparse graphs include a small number of edges so that the search space of the *OHC* is greatly reduced. For example, the *TSP* on the bounded degree graphs can be resolved in time $O((2 - \varepsilon)^n)$ [11] where $\varepsilon$ relies on the maximum degree of a vertex. In the research of approximation algorithm, Gharan and Saberi [12] proposed the polynomial-time approximation schemes based upon the bounded genus graphs where the constant factor is $22.51(1 + \frac{1}{n})$ for the planar *TSP*. For the metric *TSP* with bounded intrinsic dimension, Bartal, Gottlieb and Krauthgamer [13] have designed the randomized polynomial-time algorithm that is able to compute a $(1 + \varepsilon)$-approximation to the optimal solution where $\varepsilon > 0$. For the *TSP* on the $K_n$, there are no such good results. Whether one pursues the optimal solution or explores the approximations for the *TSP*, he or she will get the better results on the sparse graphs than those on the complete graph $K_n$. The question is how to reduce the *TSP* on the $K_n$ to the *TSP* on the sparse graphs.

This topic is the research of this paper. Given the *TSP* on the $K_n$, our objective is to eliminate the number of the irrelevant edges as most as possible and lose the number of the *OHC* edges as least as possible. After many irrelevant edges are trimmed, the sparse graphs will be obtained. If the sparse graph contains the original *OHC*, the exact or approximation algorithms will consume less time to find the *OHC* or approximations in the sparse graphs. Otherwise, if the sparse lose a few *OHC* edges, the exact or approximation algorithms will find the approximations in the sparse graphs.

To eliminate the edges out of the *OHC*, the difference between the *OHC* edges and the other edges will be explored. According to the *k−opt moves*, Hougardy and Schroeder [14] proved some edges cannot belong to the *OHC*. They presented a three-stage combinatorial algorithm to trim a lot of irrelevant edges. The experiments showed the Concorde *TSP* solver was speeded up 11 times to resolve the Euclidean *TSP* instance *d*2103 on the sparse graph.

Our work computes a sparse graph for *TSP* according to the frequency graph. In previous work [15, 16, 17], we compute the frequency graph with a set of the optimal 4-vertex paths with given endpoints where the frequency of the *OHC* edges is much higher than the average frequency of all the edges. The results benefit from the good property of these specific optimal paths which have many intersections of edges with the *OHC*. Since the frequency of the *OHC* edges are much bigger than that of most of the other edges, the minimum frequency of the *OHC* edge can be taken as the frequency threshold to eliminate the other edges with low frequency so that it is possible to compute a sparse graph for *TSP*.

In a following paper [18], Wang and Remmel computed the frequency graphs with the frequency quadrilaterals rather than the specific optimal 4-vertex paths. They listed the six frequency quadrilaterals for a weighted quadrilateral *ABCD* in the $K_n$. Based on the six frequency quadrilaterals, they formulated a binomial distribution model to derive the lower bound of the frequency of the *OHC* edge *e* as $\lfloor (\frac{7}{3} + \frac{4}{3(n-3)}) N \rfloor$ where *N* represents the number of the frequency quadrilaterals containing *e*. The probability that an *OHC* edge has the minimum frequency $\lfloor (\frac{7}{3} + \frac{4}{3(n-3)}) N \rfloor$ tends to zero for big *N* and *n*. In average case, the event that an *OHC* edge *e* has the frequency above $\lfloor (3 + \frac{2}{n-2}) N \rfloor$ has the maximum probability. It means that the frequency of the *OHC* edges will be bigger than $\lfloor (3 + \frac{2}{n-2}) N \rfloor$ in most cases. The experiments showed that the actual minimum frequency of the *OHC* edge was bigger than $\lfloor (3 + \frac{2}{n-2}) N \rfloor$ for most *TSP* instances. Moreover, the minimum frequency of the *OHC* edges increases according to *n*. Therefore, it is feasible to compute a residual graph using the frequency 3*N* as a frequency threshold.

Given the $K_n$, we first compute the corresponding frequency graph with the frequency quadrilaterals in $K_n$. After we eliminate the edges with low frequency according to the minimum frequency of the *OHC* edge, we will obtain the first preserved graph $G_1$ containing the *OHC*. A natural idea is that we can repeat the procedure for the edges in the $G_1$ if the edges in the $G_1$ are included in many quadrilaterals. That is, we compute the frequency graph of $G_1$ and trim the edges with lower frequency according to the other minimum frequency of the *OHC* edge. Furthermore, if the preconditions in the preserved graphs are sufficient, this procedure can be iterated several times until the final preserved graph is sparse enough. Based on the sparse graphs, the computation time of the algorithms for *TSP* will be greatly reduced. Once the sparse graph has the good properties, such as planarity, *k*-edge connected, bounded degree, bounded tree-width or genus, etc., the complexity of *TSP* will be lowered. A first question is how many possible edges we should eliminate at each computation cycle according to the minimum frequency of the *OHC* edge? The second question is how many cycles we can run the procedure to guarantee the preserved graphs to contain the *OHC*. Since the answers to the first question only concerns the number of the deleted edges at each computation cycle, the stop computation cycle must be given to terminate the computation procedure to output the sparse graph for *TSP*.

The outline of this paper is given as follows. In section 2, we shall briefly introduce the frequency quadrilaterals and the probability model for the *OHC* edges. A criterion is derived to eliminate how many possible edges whereas the *OHC* edges are kept intact. In section 3, we shall introduce the iterative algorithm to compute a sparse graph based on the frequency quadrilaterals. The maximum computation cycle and the stop computation cycle are also given. The iterative algorithm is tested with tens of real-world *TSP* examples in section 4. The preserved graphs in the computation process will be shown. The conclusions and possible future research are given in the last section.

## 2    The frequency quadrilaterals

The frequency quadrilateral is a kind of special frequency graph $K_i$ where $i = 4$ [18]. The frequency quadrilateral is computed with the six optimal 4-vertex paths in one corresponding quadrilateral. Here we only consider the frequency quadrilaterals derived from the general weighted quadrilaterals $K_4$. Each weighted quadrilateral just includes 6 optimal 4-vertex paths ($OP^4$) and one *OHC*. The $OP^4$s with given endpoints in a quadrilateral *ABCD* is computed as follows.

Given a quadrilateral *ABCD* in $K_n$, it includes six edges (*A*, *B*), (*A*, *C*), (*A*, *D*), (*B*, *C*), (*B*, *D*) and (*C*, *D*). The distances of the six edges are $d(A, B)$, $d(A, C)$, $d(A, D)$, $d(B, C)$, $d(B, D)$ and $d(C, D)$, respectively. Appoint two endpoints, such as *A* and *B*, there are two 4-vertex paths $P_1=(A, C, D, B)$ and $P_2=(A, D, C, B)$ containing the four vertices *A*, *B*, *C* and *D*. Their distances are computed as $d(P_1)= d(A, C) + d(C, D) + d(B, D)$ and $d(P_2)= d(A, D) + d(C, D) + d(B, C)$. We assume the distances of the two paths are unequal, i.e., $d(P_1) \ne d(P_2)$. One path must be shorter than the other one. We take the shorter path $P_1$ or $P_2$ as the $OP^4$ for the two end vertices *A* and *B*. Since we have six pairs of endpoints according to the four vertices *A*, *B*, *C* and *D*, there are six $OP^4$s in the quadrilateral *ABCD*. According to the distances of edges, the 6 $OP^4$s are computed with the four-vertex and three-line inequality array [19].

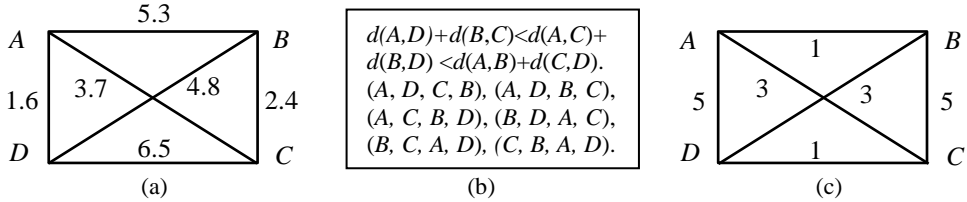

**Fig. 1.** The quadrilateral *ABCD* (a), the inequality array and the six $OP^4$s (b), and the frequency quadrilateral *ABCD* (c)

For example in Fig. 1, Fig. 1 (a) is the quadrilateral *ABCD* where the *OHC* = (*A, C, B, D, A*). Fig. 1 (b) is the inequality array $d(A, D) + d(B, C) < d(A, C) + d(B, D) < d(A, B) + d(C, D)$ according to the distances of edges and the six $OP^4$s derived based on the inequality array. Fig. 1 (c) is the frequency quadrilateral *ABCD* computed with the 6 $OP^4$s. It is clear that the 4 edges (*A, D*), (*A, C*), (*B, C*) and (*B, D*) with the top frequencies 5, 5, 3 and 3 belong to the *OHC* = (*A, C, B, D, A*).

In a frequency quadrilateral *ABCD*, the pairwise non-adjacent edges have the same frequency 5, 3 or 1. For example in Fig. 1, the edges (*A, D*)&(*B, C*) have the frequency 5, the edges (*A, C*)&(*B, D*) have the frequency 3, and the edges (*A, B*)&(*C, D*) have the frequency 1, respectively. It is the distances of the pairwise non-adjacent edges that conclude their frequencies. The bigger the summed distance of the two non-adjacent edges is, the smaller their frequency will be. For example in Fig. 1, the summed distances of the three pairs of non-adjacent edges (*A, D*)&(*B, C*), (*A, C*)&(*B, D*) and (*A, B*)&(*C, D*) are 4.0, 8.5 and 11.8, respectively. However, their frequencies are 5, 3 and 1, respectively.

The distances of the edges in a quadrilateral *ABCD* are various. According to the three summed distances of the three pairs of non-adjacent edges, they will produce six inequality arrays. Each inequality array determines a set of six $OP^4$s and a corresponding frequency quadrilateral. Thus, six distinct frequency quadrilaterals *ABCD* are computed and shown in Fig. 2 (a)-(f) [18]. The summed distance array of the quadrilateral *ABCD* is listed below the corresponding frequency quadrilateral *ABCD*.

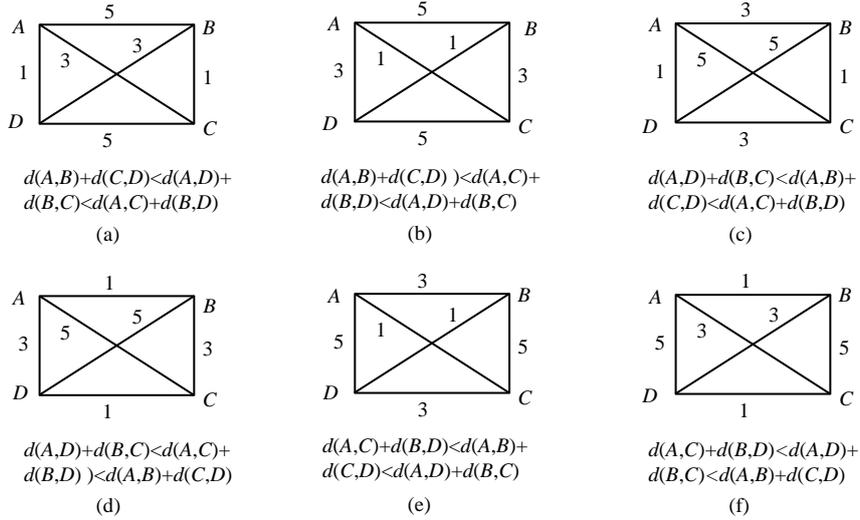

**Fig. 2.** The six frequency quadrilaterals *ABCD* for a quadrilateral *ABCD*

Although there are a number of weighted quadrilaterals, they are classified into six kinds according to the six frequency quadrilaterals in Fig.2. For each kind of quadrilaterals *ABCD*, the distances of the six edges conform to the same inequality array. In addition, their *OHC* is determined by the corresponding inequality array. Let's analyze the six frequency quadrilaterals for a quadrilateral *ABCD*. Firstly, the frequency of the six edges in each frequency quadrilateral is $f$=5, 3 and 1. The frequency of edges in the frequency quadrilaterals composes a stable frequency space {1, 3, 5}. For the three adjacent edges containing a vertex, such as *AB*, *AC* and *AD*, they have the dif-ferent frequency. Therefore, the difference between adjacent edges can be clearly characterized by their frequency. In addition, the two adjacent edges containing a vertex with the big frequency 5 and 3 are included in the *OHC* whereas another adjacent edge with the frequency 1 is not. Next, the two non-adjacent edges have the equal frequency. If one edge belongs to the *OHC*, the opposite edge must be in the *OHC* and vice versa. If we find two adjacent *OHC* edges in a quadrilateral, the other two *OHC* edges are also concluded.

Secondly, we see the frequency of an edge in the six frequency quadrilaterals. For an edge $e \in \{(A, B), (A, C), (A, D), (B, C), (B, D), (C, D)\}$, it has the frequency $f=5, 3$ and 1 twice in the six frequency quadrilaterals, respectively. If the frequency $f \in \{5, 3, 1\}$ of $e$ is taken as a random variable, we see the probability $p$ that $e$ has the frequency $f=5, 3$ and 1 is equal to 1/3 based on the six frequency quadrilaterals. We note the probability $p(f=5) = p(f=3) = p(f=1) = 1/3$ for $e$ whose frequency $f=5, 3$ and 1 in a random frequency quadrilateral containing the $e$.

Given a *TSP* with *n* vertices, there are $\binom{n}{4}$ weighted quadrilaterals. Each quadrilateral includes six edges. Thus, every edge is included in $\binom{n-2}{2}$ quadrilaterals. For an edge $e$ in each of these corresponding frequency quadrilaterals, the possible frequency of $e$ may be 5, 3 or 1. Given a random frequency quadrilateral containing $e$, we assume the frequency quadrilateral has the equal probability to be one of the six frequency quadrilaterals in Fig.2. Thus, the probability $p(f=5) = p(f=3) = p(f=1) = 1/3$ that $e$ has the frequency 5, 3 and 1 in the frequency quadrilaterals. When we compute the frequency $F(e)$ of $e$ with $N$ frequency quadrilaterals containing $e$, the frequency $F(e) = (5p(f=5) + 3p(f=3) + p(f=1))N = 3N$. It is the average frequency of all edges.

For the *OHC* edges, there are some special frequency quadrilaterals where their frequency $f = 5, 3$ rather than 1. In the $K_n$, each two adjacent *OHC* edges are included in $n-3$ quadrilaterals and each two opposite *OHC* edges are contained in a quadrilateral. In the corresponding frequency quadrilateral containing two opposite *OHC* edges, the frequency of the two *OHC* edges are 5 or 3. Otherwise, the two opposite *OHC* edges will be replaced by the other two non-adjacent edges in the quadrilateral. According to the observations, Wang and Remmel [18] constructed the *n*-3 frequency quadrilaterals for an *OHC* edge $e$ where its frequency is 3 or 5. In the rest frequency quadrilaterals, they assume the frequency 1, 3 and 5 of $e$ has the equal probability 1/3. Thus, they gave the probability that $e \in OHC$ has the frequency 1, 3 and 5 in a frequency quadrilateral as $p(f=5) = p(f=3) = \frac{1}{3} + \frac{1}{3(n-2)}$ and $p(f=1) = \frac{1}{3} - \frac{2}{3(n-2)}$. When we compute the frequency of $e \in OHC$ with $N$ random frequency quadrilaterals containing $e$, the expected frequency $F(e) = 3N + \frac{2N}{n-2}$. One sees the expected frequency of an edge $e \in OHC$ computed with $N$ random frequency quadrilaterals is bigger than the expected frequency $3N$ of a common edge.

According to the probability $p(f=5) = p(f=3) = \frac{1}{3} + \frac{1}{3(n-2)}$ and $p(f=1) = \frac{1}{3} - \frac{2}{3(n-2)}$, we know that the probability $p(f=3,5) > \frac{2}{3}$ for an *OHC* edge $e$ in a frequency quadrilateral in $K_n$. When we compute the frequency of an edge with $N$ frequency quadrilaterals containing $e$, the total frequency of an *OHC* edge will be bigger than $3N$. Certainly, the average frequency $\bar{f}(e)$ of the *OHC* edge $e$ will be bigger than 3. Therefore, it is necessary to consider the edges $e$ with the frequency $F(e) > 3N$ or $\bar{f}(e) > 3$ for *TSP*. We are interested in how many edges with the frequency $F(e) > 3N$ or $\bar{f}(e) > 3$ when each of them is computed with $N$ frequency quadrilaterals.

Given an edge $e$ in the six frequency quadrilaterals in Fig.2, there are four frequency quadrilaterals where $e$ has the frequency 3 and 5. Thus, the probability $p(f≥3)$ of the case $f≥3$ for an edge $e$ is equal to 2/3, i.e., $p(f≥3)= 2/3$. It says the $e$ has the probability 2/3 that its frequency is bigger than the expected frequency 3 in a random frequency quadrilateral. When we choose $N$ random frequency quadrilaterals containing $e$, there will be $\left[\frac{2N}{3}\right]$ frequency quadrilaterals where the frequency is above 3 and $\left[\frac{N}{3}\right]$ frequency quadrilaterals where the frequency is below 3. If we take 3 as the frequency threshold to eliminate the edges with smaller frequency, the edge $e$ will be preserved $\left[\frac{2N}{3}\right]$ times. Thus, the probability that $e$ is preserved is $\frac{2}{3}$ when we take $f=3$ as the frequency threshold to eliminate the edges with lower frequency. In this case, the total frequency $F(e) = 3N$ and average frequency $\bar{f}(e) = 3$. If we take the total frequency $F = 3N$ or average frequency $\bar{f} = 3$ as the frequency threshold, the edges $e$ with the probability $p(f≥3) ≥2/3$ in each frequency quadrilateral will be preserved. Considering the $\binom{n}{2}$ edges in the $K_n$, we will preserve at most $\frac{2}{3}\binom{n}{2}$ edges with the big frequency for *TSP*. Since the *OHC* edge $e$ has the big frequency $F(e) > 3N$ or $\bar{f}(e) > 3$, the *OHC* edges will be preserved with a big probability. In fact, an *OHC* edge $e$ has the probability $p(f≥3) ≥2/3$ in each frequency quadrilateral so it is preserved.

It mentions that we will preserve some edges with the probability $p(f≥3)<2/3$ when we take $F = 3N$ or $\bar{f} = 3$ as the frequency threshold. For example, the edges have the probability $p(f=5)>1/2$ and $p(f=3) + p(f=5) < 2/3$. Thus, more irrelevant edges with the big frequency $F(e) > 3N$ and average frequency $\bar{f}(e) > 3$ will be preserved. If the edges in the preserved graph are contained in many frequency quadrilaterals, it is necessary to iterate the computation process to eliminate these edges in the next computation cycles. In the following section, we will give the iterative algorithm to trim these edges step by step until a sparse graph for *TSP* is computed.

## 3    The iterative algorithm

We fist give the iterative algorithm and then discuss the stop computation cycle. Given an original graph $G_0(V_0, E_0)$, $/V_0/ = n$ and $/E_0/ =\binom{n}{2}$ in general. When the frequency of every edge is computed with $N$ frequency quadrilaterals in $G_0(V_0, E_0)$, we choose $\left[\frac{2}{3}|E_0|\right]$ edges with top frequency to compose a second graph $G_1(V_0, E_1)$ for *TSP* according to the probability $p(f = 5) = p(f = 3) = \frac{1}{3} + \frac{1}{3(n-2)}$ and $p(f = 1) = \frac{1}{3} - \frac{2}{3(n-2)}$. After that, if $N$ is big for the edges in $G_1(V_0, E_1)$, we can use the same method to compute a third graph $G_2(V_0, E_2)$ with an even smaller number of edges where $/E_2/ = \left[\left(\frac{2}{3}\right)^2 |E_0|\right]$. Furthermore, if $N$ is still big for the edges in $G_k(V_0, E_k)$ where $k > 2$, we can keep executing the computation until a sparse graph containing $\left[\left(\frac{2}{3}\right)^k |E_0|\right]$ edges is generated. We expect the final sparse graph to include $O(nlog_2n)$ edges so that the better polynomial-time algorithms or polynomial-time approximate schemes are designed for *TSP* based on these sparse graphs. It notes that *N* should be sufficiently big for the edges in $G_k(V_0, E_k)$ at each computation cycle where $k \geq 0$. Otherwise, the probability $p_5(e)$ and $p_3(e)$ of the *OHC* edges will be much smaller than the $p(f = 5) = p(f = 3) = \frac{1}{3} + \frac{1}{3(n-2)}$ and smaller than that of the other edges. One will wonder the probability $p(f = 5)$, $p(f = 3)$ of the *OHC* edges will become smaller in the preserved graphs and they will be eliminated in the computation process. We will explore the change of the probability $p(f = 5)$, $p(f = 3)$ and $p(f = 1)$ for the *OHC* edges in the computation process in another paper. Here we will focus on the algorithm to compute the sparse graphs for *TSP*. In the next section, we will see the *OHC* edges usually have the bigger average frequency than that of the $1/3|E_k|$ edges to be eliminated in the preserved graphs for the *TSP* instances. Under the assumption, we give the iterative algorithm in Table (2).

**Table 1.** The iterative algorithm to compute a sparse graph for *TSP*

| Step | The pseudo codes of the iterative algorithm |
|------|---------------------------------------------|
| 1 | Input the initial graph $G_k(V_0, E_k)$ where $k := 0$, $/V_0/ = n$ and $/E_0/ = \binom{n}{2}$. |
| 2 | Do{ |
| 3 | For each edge $e \in E_k$ { Compute the average frequency $\bar{f}(e)$ of $e$ with $N$ frequency quadrilaterals containing $e$.} |
| 4 | Order the $/E_k/$ edges according to their $\bar{f}(e)$s from big to small values. |
| 5 | Preserve the previous $\left[\frac{2}{3}|E_k|\right]$ edges to the next graph $G_{k+1}(V_0, E_{k+1})$. |
| 6 | Replace $G_k(V_0, E_k)$ with $G_{k+1}(V_0, E_{k+1})$ and assign $k := k + 1$. |
| 7 | }While($/E_{k+1}/ \geq cn$). |
| 8 | Output the sparse graph with $cn$ edges. |

The first step is to input an initial weighted graph $G_0(V_0, E_0)$ with $n$ vertices and $/E_0/$ edges. Generally, $/V_0/ = n$ and $/E_0/ = \binom{n}{2}$. Assign the initial value of computation cycle $k = 0$. In the following steps, the iterative algorithm generates a sparse graph with less than $cn$ edges where $c \approx [log_2 n]$. At the $k^{th}$ computation cycle where $k \geq 0$, the algorithm begins with a input graph $G_k(V_0, E_k)$ and outputs the next preserved graph $G_{k+1}(V_0, E_{k+1})$ with $/E_{k+1}/= \left[\left(\frac{2}{3}\right)^{k+1} |E_0|\right]$ edges according to the average frequency of edges in the $G_k(V_0, E_k)$. We use the average frequency of edges instead of their total frequency to avoid bias to some edges since the edges will be contained in different number of frequency quadrilaterals in the in $G_k(V_0, E_k)$. The average frequency $\bar{f}(e)$ of every edge $e$ in $G_k(V_0, E_k)$ is computed as follows. Firstly, the $N$ quadrilaterals containing the edge $e$ are chosen and the frequency $f(e)$ of $e$ is enumerated from the 6 $OP^4$s in each of the quadrilaterals. Given the frequency of $e$ is $f_j(1 \leq j \leq N)$ in the $j^{th}$ frequency quadrilateral, the average frequency of $e$ is computed as $\bar{f}(e) = \frac{1}{N}\sum_{j=1}^{N} f_j$. In the iterative computation process, the number of the frequency quadrilaterals containing every edge in $G_k(V_0, E_k)$ will not be equal. It is fair to compare the average frequency of edges rather than their total frequency in $G_k(V_0, E_k)$. Therefore, it is rational to keep the $\left[\frac{2}{3}|E_k|\right]$ edges with the big average frequency for *TSP*.

After the average frequency $\bar{f}(e)$s of the $|E_k|$ edges are computed, we order them from big to small values to form a frequency sequence at step 4. Given the average frequency of the $j^{th}$ edge is $\bar{f}_j (1 \leq j \leq |E_k|)$, we note the frequency sequence as $(\bar{f}_1, \bar{f}_2, \cdots, \bar{f}_{|E_k|})$ where $\bar{f}_1$ is the maximum frequency and $\bar{f}_{|E_k|}$ is the minimum frequency. At the $5^{th}$ step, the previous $\left[\frac{2}{3}|E_k|\right]$ edges are chosen to compose the next graph $G_{k+1}(V_0, E_{k+1})$. To continue the recursive computations, we replace the graph $G_k(V_0, E_k)$ with the graph $G_{k+1}(V_0, E_{k+1})$ and assign $k := k + 1$. That is, the edge set $E_k$ in the $G_k$ is substituted with the edge set $E_{k+1}$ in $G_{k+1}$. The iterative algorithms will always be executed until the terminal condition $|E_{k+1}| \leq cn$ is met. At last, it outputs the final sparse graph with less than $cn$ edges.

When we run the iterative algorithm based on the frequency quadrilaterals, the preserved graphs will have the smaller and smaller number of edges according to the computation cycle $k$. Given through $k$ iterations, we will obtain a sparse graph with $cn$ edges. The maximum iterations $k_{max}$ is given as formula (1). Many researchers take the graphs with $O(n\log_2 n)$ edges as the sparse graphs. The number $O(n\log_2 n)$ increases nearly in a linear way according to $n$. Thus, we take $c = \log_2 n$ for general *TSP* instances. The maximum computation cycle becomes $k_{max} = \left\lfloor log_{\frac{2}{3}}\left(\frac{2 log_2 n}{n-1}\right)\right\rfloor$. At the $k_{max}^{th}$ computation cycle, the iterative algorithm will output a sparse graph with $[n\log_2 n]$ edges for general *TSP*.

$$k_{max} = \left\lfloor log_{\frac{2}{3}}\left(\frac{2c}{n-1}\right)\right\rfloor \qquad (1)$$

If every $K_4$ includes only one *OHC* and the six $OP^4$s, the average frequency of the *OHC* edges will be bigger than the expected frequency 3 whereas the average frequency some of the other edges will be below 3. In this case, we can eliminate $\frac{1}{3}$ edges with small average frequency. This is the theoretical case. For real-world instances, some $K_4$s in the $K_n$ include more than six $OP^4$s as they contain the equal-weight edges. The selection of the right 6 $OP^4$s becomes hard to compute a unique frequency quadrilateral. If the wrong $OP^4$s in these $K_4$s are used, the average frequency of some *OHC* edges in the $K_n$ will become smaller. These *OHC* edges will be eliminated with a big probability. For general *TSP*, the iterative algorithm will work well to compute a sparse graph for *TSP*. Even though for the special *TSP* examples, the experiments showed that the iterative algorithm still works if we add the random small distances to the edges' distances in advance [18].

We are interested in how many *OHC* edges will be lost in the computation process. At the $(k + 1)^{th}$ computation cycle, we will maintain $\left[\frac{2}{3}|E_k|\right]$ edges in the preserved graph. In other words, we will throw away $\left[\frac{1}{3}|E_k|\right]$ edges according to the average frequency of edges. In the graph $G_k(V_0, E_k)$, the probability that an edge $e$ is abandoned is $p(e \notin E_{k+1}) = 1/3$. The *OHC* includes $n$ edges. If we do not consider the frequency of edges, the probability that an edge $e \in OHC$ in $G_k(V_0, E_k)$ is $p(e \in OHC) =$ $\frac{n}{|E_k|}$ where $|E_0| = \binom{n}{2}$. At the $k^{th}$ ($k \geq 1$) computation cycle, the number of the edges in the input graph $G_{k-1}(V_0, E_{k-1})$ is $\left[\left(\frac{2}{3}\right)^{k-1} |E_0|\right]$. We assume the graph $G_{k-1}(V_0, E_{k-1})$ includes the *OHC* so that the probability of an edge $e \in OHC$ is $p(e \in OHC)$ $= \frac{n}{\left(\frac{2}{3}\right)^{k-1} |E_0|}$. Every edge has the probability 1/3 to be abandoned at the $k^{th}$ computation cycle. Thus, the probability that the edge $e \in OHC$ is abandoned in the next graph $G_k(V_0, E_k)$ is computed as formula (2).

$$p(e \in OHC \wedge e \notin E_k) = \frac{1}{3} \times \frac{n}{\left(\frac{2}{3}\right)^{k-1} |E_0|} \tag{2}$$

If one or more *OHC* edges are lost at the $k^{th}$ computation cycle, we have $np(e \in OHC \wedge e \notin E_k) \geq 1$. Therefore, the formula (4) holds if $|E_0| = \binom{n}{2}$.

$$k \geq 2 + \left\lfloor \log_{\frac{2}{3}} \left(\frac{n}{n-1}\right) \right\rfloor \tag{3}$$

As $n \to \infty$, $k \geq 2$. It means the iterative algorithm can be executed at least 2 times without losing the *OHC* edges. If $m$ *OHC* edges are lost where $m$ is a small constant, we can derive the computation cycles $k \geq 2 + \left\lfloor \log_{\frac{2}{3}} \left(\frac{1}{m}\right) \right\rfloor$. The computation cycle becomes bigger as $m$ rises. It means that we can run the algorithm more times to compute a sparse graph on condition that only a small number of the *OHC* edges are lost. For example, if $m = \frac{9}{4}$, we can run the algorithm at least $k = 4$ times. However, we may lose only $m < 3$ *OHC* edges.

In fact, the formula (2) is the average probability for an arbitrary edge $e$ in any given $n$ edges in $G_{k-1}(V_0, E_{k-1})$. In our algorithm, we maintain the edges according to the frequency of edges rather than the random selection. For each of the edges in the *OHC*, their average frequency computed with the $N$ frequency quadrilaterals in $G_{k-1}(V_0, E_{k-1})$ will be bigger than the average frequency of all of the edges. In the frequency sequence $(\bar{f}_1, \bar{f}_2, \cdots, \bar{f}_{|E_{k-1}|})$, if the number of edges with the average frequency below the 3 is bigger than $\left[\frac{1}{3}|E_{k-1}|\right]$, the probability that the *OHC* edges will be maintained in $G_k(V_0, E_k)$ tends to 1 but not 2/3. Moreover, the probability that an *OHC* edge is neglected in $G_k(V_0, E_k)$ is small than 1/3 based on the frequency quadrilaterals. Therefore, the edges in the *OHC* have a much bigger probability that they will be maintained to the $G_k(V_0, E_k)$ as $N$ is big enough. According to the average frequency of edges, the probability $p(e \in OHC \wedge e \notin E_k)$ will be much smaller than that computed with the formula (2) based on the random selection. Thus, we can run the iterative algorithm more times than that restricted by formula (3) for general *TSP*. Meanwhile, we will obtain an even sparser graph containing the *OHC* for *TSP*.

Many incomplete quadrilaterals will appear in the computation process because the $\left[\frac{2}{3}|E_{k-1}|\right]$ edges with small frequency are abandoned at the $k^{th}$ computation cycle. These incomplete quadrilaterals contain less than 6 edges as well as less than 6 $OP^4$'s. If these incomplete quadrilaterals are used to compute the frequency of edges, the average frequency of edges will not equal 3. The probability $p(f = 5) = p(f = 3) = p(f = 1) = \frac{1}{3}$ will not be right based on these incomplete frequency graphs. The probability that an edge $e$ has the frequency above 3 is not equal to $\frac{2}{3}$ in the various incomplete frequency quadrilaterals. Therefore, we should use a different ratio rather than $\frac{1}{3}$ to discard the number of edges according to their average frequency, especially in the later computation stage.

In the later computation process, most of edges in the preserved graphs will only be included in the incomplete quadrilaterals. According to various incomplete quadrilaterals, it is hard to find a suitable ratio to delete the proper number of edges in $G_k(V_0, E_k)$. If we use the constant ratio $\frac{1}{3}$ to abandon the edges with small frequency, some or many *OHC* edges will be neglected, too. To guarantee the *OHC* edges in the last preserved graph, we will find the stop computation cycle $k_s$ to terminate the iterative algorithm.

If $N$ is big enough for an edge $e \in OHC$ in $G_k(V_0, E_k)$, the average frequency of $e$ will be bigger than the average frequency of the total $|E_k|$ edges based on frequency quadrilaterals. Thus, the average frequency of $e \in OHC$ will be bigger than 3 when it is computed with $N$ frequency quadrilaterals. We enumerate the number of edges with the average frequency less than 3 and note it as $N_{<\bar{f}}$. When we use the constant ratio $\frac{1}{3}$ to trim the $\frac{1}{3}|E_k|$ edges with small frequency, the *OHC* will be maintained in the preserved graph if $N_{<\bar{f}} > \frac{1}{3}|E_k|$. It means we just eliminate the $\frac{1}{3}|E_k|$ edges whose average frequency is below 3. However, we will eliminate some *OHC* edges if we meet $N_{<\bar{f}} \leq \frac{1}{3}|E_k|$ in the graph $G_k(V_0, E_k)$. Therefore, the inequality $N_{<\bar{f}} \leq \frac{1}{3}|E_k|$ is taken as the restriction to determine the stop computation cycle $k_s$ and terminate the iterative algorithm. Given a *TSP*, the iterative algorithm can always run until it reaches the stop computation cycle. Once $N_{<\bar{f}} \leq \frac{1}{3}|E_k|$, we should be careful to implement the iterative computation. Some *OHC* edges whose average frequency is above but near to the expected frequency 3 will be abandoned at this computation cycle.

In the computation process, the number of edges with the average frequency below 3 will become less and less according to $k$. On the other hand, the number of edges with average frequency above 3 will become relatively bigger according to $k$. Therefore, we can always find such a preserved graph $G_k(V_0, E_k)$ where $N_{<\bar{f}} \leq \frac{1}{3}|E_k|$. If the minimum average frequency of the *OHC* edge is still bigger than 3 or the average frequency of the *OHC* edges is further beyond that of the $\frac{1}{3}|E_k|$ edges with small frequency, we may proceed the iterative algorithm one or a few more times even though the $N_{<\bar{f}} \leq \frac{1}{3}|E_k|$. In this case, the computation cycle $k$ will be close to $k_{max}$ and the residual graph will be very sparse. However, we cannot guarantee to preserve all of the *OHC* edges in the following preserved graphs for the worst cases of *TSP* once $k > k_s$. In the actual computation process, we enumerate the edges with the average frequency below the expected frequency 3 simultaneously. Once $N_{<\bar{f}} \leq \frac{1}{3}|E_k|$, it means that we may throw away some *OHC* edges at this computation cycle. It is the time to stop the iterative algorithm and take the output graph with $|E_{k-1}|$ edges for *TSP*.

It mentions that many incomplete quadrilaterals will be generated in the computation process. Fig. 3 (a) and (b) shows two kinds of incomplete quadrilaterals we consider in our algorithm, especially at the final stages of the algorithm. Their possible frequency quadrilaterals (1), (2) are shown on their right sides. These incomplete frequency quadrilaterals are computed with the $OP^4$s in the two incomplete quadrilaterals. In the frequency quadrilaterals (1) and (2), the numbers on the edges are their frequency. Obviously, the probability 4/5 and 1 that we preserve the edges based on the two incomplete frequency quadrilaterals is bigger than 2/3 according to frequency quadrilaterals. It means we should throw away $\frac{1}{5}|E_k|$ and 0 edges according to their frequency or average frequency computed with the two kinds of incomplete frequency quadrilaterals. It suggests us to keep more edges in the computation process when we use a lot of such incomplete frequency quadrilaterals. In the later computation cycles, we will have many such incomplete quadrilaterals. In this case, we generally cannot use the constant ratio $\frac{1}{3}$ to eliminate the $\frac{1}{3}|E_k|$ edges with low frequency to compute the next sparse graph for *TSP*.

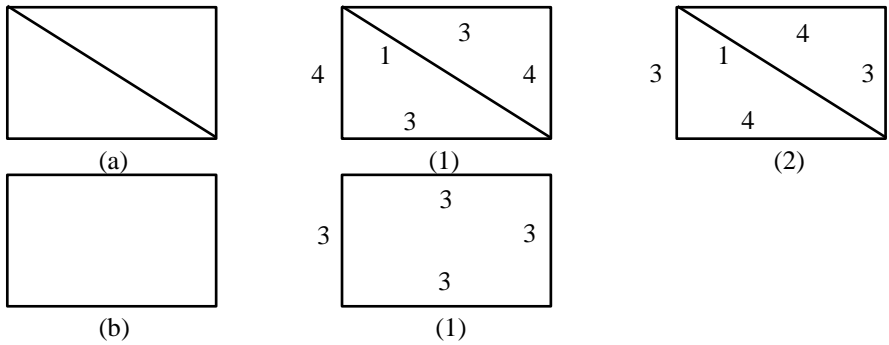

**Fig. 3.** Two kinds of incomplete quadrilaterals and their corresponding frequency graphs

## 4      The experiments and analysis

We use some *TSP* examples in *TSPLIB* [20] to verify the performance of the iterative algorithm. Four families of *TSP* instances are used to test the iterative algorithm. They are the *Euclidean*, *GEO*, *ATT* and *Matrix TSP* in *TSPLIB*. The frequency quadrilaterals are used to compute the frequency of every edge. At each computation cycle, we use all of the frequency quadrilaterals containing every edge in the input graph to compute their frequency. During the computation, the number of the frequency quadrilaterals containing every edge will not be equal due to the omission of edges. In this case, the bias will occur to the edges if we keep them according to their total frequency. To prevent such bias, the average frequency of every edge is computed for comparison and the edges with big average frequency will be maintained. For these instances, we use the Online Concorde [21] to compute their *OHC*s first. The *OHC* is used to compute the number of the lost *OHC* edges in the preserved graphs at each computation cycle.

Besides the $K_4$s, the two kinds of incomplete quadrilaterals shown in Fig. 3 are also used to compute the frequency of edges. However, these incomplete quadrilaterals are not used until the computation cycle $k \geq \left\lfloor \frac{2}{3} log_{\frac{2}{3}}\left(\frac{2}{n-1}\right) \right\rfloor$. Before the $\left\lfloor \frac{2}{3} log_{\frac{2}{3}}\left(\frac{2}{n-1}\right) \right\rfloor$ computation cycles, most of the preserved edges are included in many $K_4$s, especially for the *OHC* edges. On the other hand, the preserved edges will be included in many incomplete quadrilaterals after $\left\lfloor \frac{2}{3} log_{\frac{2}{3}}\left(\frac{2}{n-1}\right) \right\rfloor$ computation cycles. In this case, these incomplete frequency quadrilaterals are used to compute the average frequency of the preserved edges. Since the average frequency of an edge is computed with the different numbers of complete and incomplete frequency quadrilaterals, the probability that we preserve the edges are changing in the computation process. To simplify the computation, we preserve the $\frac{2}{3}|E_k|$ edges with the big average frequency at each computation cycle. Although the incomplete frequency quadrilaterals are used, the average frequency of the *OHC* edges will still be bigger than that of the edges to be eliminated in most of the preserved graphs. Thus, the iterative algorithm can be executed several more times and the even sparser graphs will be computed.

In the experiments, we execute the algorithm $k_{max} = \left\lfloor log_{\frac{2}{3}}\left(\frac{2}{n-1}\right) \right\rfloor$ times for each *TSP* instance where $c=1$. Some vertices will be isolated in the computation process. For an isolated vertex, we add the two associative edges with the maximum average frequency in the last computation cycle to guarantee the connectivity of the preserved graphs. Thus, the number of the edges in the preserved graphs does not decrease accurately in proportion to the factor 2/3 according to $k$ in the last computation stages.

The results are shown in Tables 2, 3 and 4, respectively. The first column is the *TSP* name and the number in the name is the scale (or size) $n$ of *TSP*. For each *TSP* instance, we recorded the number of the preserved edges and the number of the lost *OHC* edges according to the computation cycles $k$. The number of the lost *OHC* edges is noted behind the number of edges in the preserved graphs. There is a slash between the two numbers. We only illustrate the results where the number of the lost *OHC* edges is less than 13 for the examples in the Tables 2 and 3 according to the computation cycles $k$. In the Table 4, we show the preserved graphs where more *OHC* edges are lost for the big size of *TSP* instances. Since the preserved graphs in the first several computation cycles contain the *OHC*, we give the results from the 3$^{rd}$ computation cycle for the small size of instances in Table 2, from the 5$^{th}$ computation cycle for the medium size of instances in Table 3 and from the 8$^{th}$ computation cycle for the big size of instances in Table 4. In the last, we use the preserved graphs losing at most one *OHC* edge in Tables 2, 3 and the preserved graphs losing at most 2 *OHC* edges in Table 4 to compute the parameter $c = \frac{|E_k|}{n}$ for every example. The average degree of each sparse graph is computed as $d = \left\lceil \frac{2|E_k|}{n} \right\rceil$ with the same preserved graph.

**Table 2.** The computational results of some TSP instances: 1st Table.

| TSP\k | 3 | 4 | 5 | 6 | 7 | 8 | 9 | c | d |
|---|---|---|---|---|---|---|---|---|---|
| gr17 | 43 | 30/1 | 23/5 | | | | | 1.765 | 4 |
| gr21 | 64 | 43 | 32/1 | | | | | 1.524 | 3 |
| gr24 | 83 | 57/1 | 54/1 | | | | | 2.25 | 5 |
| fri26 | 97 | 65 | 45 | 37/2 | | | | 1.730 | 4 |
| bayg29 | 121 | 81/1 | 56/2 | 44/6 | | | | 2.793 | 6 |
| dantzig42 | 257 | 172 | 115 | 81/4 | | | | 2.738 | 6 |
| att48 | 338 | 226 | 152 | 103/3 | 79/10 | | | 3.167 | 6 |
| gr48 | 335 | 221 | 148 | 99/2 | 76/8 | | | 3.1 | 6 |
| hk48 | 336 | 224 | 150 | 102/2 | 73/10 | | | 3.125 | 6 |
| eil51 | 382 | 256 | 172 | 116 | 90 | | | 1.765 | 4 |
| berlin52 | 396 | 267 | 180 | 123 | 92/6 | | | 2.365 | 5 |
| brazil58 | 491 | 325 | 217 | 146 | | | | 2.517 | 5 |
| st70 | 717 | 479 | 320 | 215/2 | 149/6 | | | 4.571 | 9 |
| eil76 | 848 | 568 | 380 | 255 | 173/2 | 128/9 | | 3.355 | 7 |
| pr76 | 848 | 566 | 379/1 | 254/3 | 172/5 | 126/12 | | 4.987 | 10 |
| gr96 | 1355 | 905 | 606 | 406 | 277 | 187/4 | 176/10 | 2.885 | 6 |
| rat99 | 1441 | 962 | 644 | 431 | 291/1 | 198/7 | | 2.939 | 6 |
| kroA100 | 1470 | 982 | 656 | 439 | 294 | 210/3 | 159/9 | 2.940 | 6 |
| kroB100 | 1470 | 982 | 656 | 439 | 294/1 | 208/3 | 194/7 | 2.940 | 6 |
| kroC100 | 1470 | 982 | 656 | 439 | 295 | 210/5 | 158/10 | 2.950 | 6 |
| kroD100 | 1470 | 982 | 656 | 439 | 296 | 204/7 | 198/8 | 2.960 | 6 |
| rd100 | 1470 | 982 | 657 | 440 | 295/5 | 204/10 | | 4.40 | 9 |
| eil101 | 1500 | 1002 | 670 | 448 | 300 | 202/4 | | 2.970 | 6 |
| lin105 | 1622 | 1082 | 725 | 485/1 | 326/3 | 325/5 | 167/9 | 4.619 | 9 |
| pr107 | 1684 | 1122 | 748 | 500 | 336/3 | 232/8 | | 4.673 | 9 |
| gr120 | 2117 | 1411 | 939 | 627 | 419 | 287/7 | | 3.491 | 7 |
| pr124 | 2263 | 1510 | 1008/1 | 676/1 | 452/1 | 307/3 | 216/9 | 3.645 | 7 |
| bier127 | 2374 | 1584 | 1058 | 707 | 474 | 320/1 | 224/12 | 3.732 | 7 |
| ch130 | 2488 | 1660 | 1109 | 741 | 496 | 333/3 | 249/13 | 3.815 | 8 |
| pr136 | 2724 | 1818 | 1215 | 812 | 543 | 367 | 263/11 | 2.699 | 5 |
| gr137 | 2764 | 1844 | 1231 | 822 | 550 | 373/2 | 259/7 | 4.015 | 8 |
| pr144 | 3054 | 2038 | 1360 | 907 | 606 | 409/4 | 287/12 | 4.208 | 8 |
| ch150 | 3315 | 2212 | 1476 | 985 | 658 | 440 | 311/10 | 2.933 | 6 |
| kroA150 | 3315 | 2212 | 1478 | 988 | 660 | 442/1 | 314/5 | 2.947 | 6 |
| kroB150 | 3315 | 2212 | 1476 | 986 | 659 | 442/4 | 311/10 | 4.393 | 9 |
| Pr152 | 3404 | 2271 | 1514 | 1010/1 | 675/3 | 454/3 | 314/10 | 6.645 | 13 |
| u159 | 3726 | 2486 | 1659 | 1108 4 | 740 | 495/2 | 335/5 | 4.65 | 9 |
| si175 | 4515 | 3011 | 1996 | 1300 | 869/4 | | | 7.429 | 15 |
| brg180 | 4777 | 3186/1 | 2124/5 | | | | | 17.70 | 35 |
| rat195 | 5608 | 3740 | 2494 | 1668 | 1114 | 745 | 504/1 | 2.585 | 5 |
| d198 | 5782 | 3856 | 2572 | 1718 | 1147/1 | 766/1 | 513/11 | 3.869 | 8 |

**Table 3.** The computational results of some TSP instances: 2$^{nd}$ Table.

| TSP\k   | 5     | 6     | 7      | 8      | 9      | 10     | 11      | c      | d  |
|---------|-------|-------|--------|--------|--------|--------|---------|--------|----|
| kroA200 | 2625  | 1752  | 1170   | 783    | 527/5  |        |         | 3.915  | 8  |
| kroB200 | 2625  | 1752  | 1170   | 782/1  | 536/5  |        |         | 3.910  | 8  |
| gr202   | 2678  | 1787  | 1193   | 802    | 539/3  |        |         | 3.970  | 8  |
| ts225   | 3323  | 2217  | 1480   | 988/1  | 662/4  |        |         | 4.391  | 9  |
| tsp225  | 3323  | 2219  | 1481   | 990    | 666/1  |        |         | 2.960  | 6  |
| pr226   | 3352  | 2228  | 1487   | 993/3  | 669/4  |        |         | 6.580  | 13 |
| gr229   | 3442  | 2296  | 1532   | 1023   | 685/3  | 475/11 |         | 4.467  | 9  |
| pr264   | 4576  | 3053  | 2038/1 | 1360/1 | 913/6  | 614/10 |         | 5.152  | 10 |
| a280    | 5148  | 3434  | 2291   | 1529   | 1021/1 | 693/8  |         | 3.646  | 7  |
| pr299   | 5871  | 3916  | 2614   | 1745   | 1165/3 | 786/11 |         | 5.836  | 12 |
| lin318  | 6642  | 4430  | 2955   | 1972   | 1317/3 | 882/11 |         | 6.201  | 12 |
| rd400   | 10513 | 7010  | 4676   | 3119   | 2081/2 | 1392/4 |         | 7.798  | 16 |
| fl417   | 11426 | 7617  | 5077   | 3387   | 2260   | 1508/8 |         | 5.420  | 11 |
| gr431   | 12207 | 8140  | 5428   | 3620   | 2416   | 1615/1 | 1085/9  | 3.747  | 7  |
| pr439   | 12665 | 8445  | 5633   | 3757   | 2506/1 | 1684/2 | 1126/10 | 4.683  | 9  |
| pcb442  | 12838 | 8560  | 5708   | 3807   | 2541   | 1697/3 | 1160/9  | 5.749  | 11 |
| d493    | 15975 | 10652 | 7103   | 4737   | 3161   | 2111/1 | 1418/12 | 4.282  | 9  |
| att532  | 18604 | 12404 | 8271   | 5513   | 3677   | 2452/1 | 1646/8  | 4.609  | 9  |
| ail535  | 18813 | 12543 | 8363   | 5576   | 3718   | 2480/3 |         | 6.949  | 13 |
| si535   | 18788 | 12455 | 8180   | 5309/6 | 3542/12|        |         | 15.290 | 31 |
| pa561   | 20688 | 13791 | 9192   | 6124   | 4084   | 2724/2 | 1829/11 | 7.280  | 15 |
| u574    | 21660 | 14442 | 9630   | 6419   | 4281   | 2855/1 | 1908/11 | 4.974  | 10 |
| rat575  | 21736 | 14492 | 9663   | 6444   | 4298   | 2867   | 1913/2  | 4.986  | 10 |
| p654    | 28123 | 18748 | 12498  | 8332   | 5556   | 3706/5 | 2472/13 | 8.495  | 17 |
| d657    | 28382 | 18923 | 12617  | 8412   | 5610   | 3743   | 2497/7  | 5.697  | 11 |
| gr666   | 29166 | 19446 | 12966  | 8651   | 5770   | 3848/1 | 2568/4  | 5.778  | 12 |

We observed the results from the experimental datum: (1) A sparse graph with $cn$ edges is computed in polynomial-time where $c \ll n$. For most *TSP* instances, $c < \log_2(n)$ except for a few examples. The average degree of vertices $d = O(\log_2(n))$ in these residual graphs. (2) For most of the examples, the sparse graphs will lose a few *OHC* edges when $k$ is close to $\left\lfloor \log_{\frac{2}{3}}\left(\frac{2}{n-1}\right) \right\rfloor$, such as $k = \left\lfloor \log_{\frac{2}{3}}\left(\frac{2}{n-1}\right) \right\rfloor - 2$. At this time, the residual graphs have become very sparse and most of the preserved edges is only included in a few incomplete quadrilaterals. Therefore, most of *TSP* instances are reduced to a *TSP* on the sparse graphs. (3) The number of edges in the preserved graphs decreases in proportion to the factor 2/3 according to computation cycles $k$. The computation cycle $k$ is less than $\left\lfloor \log_{\frac{2}{3}}\left(\frac{2}{n-1}\right) \right\rfloor$ to compute a sparse graph with $O(n\log_2(n))$ edges. (4) All the preserved graphs include the *OHC* for these examples as $k \leq 3$. When $k = 4$, only the brg180 begins losing the *OHC* edges whereas the preserved graphs of the other examples still include the *OHC*. (5) In the last two computation cycles when $k \geq \left\lfloor \log_{\frac{2}{3}}\left(\frac{2}{n-1}\right) \right\rfloor - 2$, quite a few *OHC* edges are lost in the pre-served graphs (not shown in the Tables 2, 3 and 4) because the preserved edges are seldom included in the complete quadrilaterals. Thus, we cannot eliminate the edges according to their frequency since most of them are zero. We generally cannot compute a residual graph with *n* edges with the iterative algorithm.

Table 4. The computational results of some TSP instances: 3$^{rd}$ Table.

| TSP\k   | 8      | 9     | 10      | 11      | 12      | 13       | 14     | c      | d  |
|---------|--------|-------|---------|---------|---------|----------|--------|--------|----|
| u724    | 10216  | 6812  | 4544    | 3031/1  | 2054/8  |          |        | 4.186  | 8  |
| rat783  | 11950  | 7968  | 5314    | 3547    | 2378/6  |          |        | 4.530  | 9  |
| pr1002  | 19569  | 13047 | 8699    | 5800/1  | 3868/4  |          |        | 5.788  | 12 |
| u1060   | 21903  | 14603 | 9736    | 6491    | 4335/8  |          |        | 6.124  | 12 |
| vm1084  | 22903  | 15269 | 10180   | 6787/2  | 4526/7  |          |        | 6.261  | 13 |
| pcb1173 | 26822  | 17882 | 11922   | 7949/1  | 5300/2  | 3538/13  |        | 4.518  | 9  |
| d1291   | 32491  | 21659 | 14440/1 | 9627/1  | 6422/3  | 4338/10  |        | 7.457  | 15 |
| rl1304  | 33150  | 22100 | 14734/2 | 9825/9  | 6551/22 |          |        | 11.316 | 23 |
| rl1323  | 33150  | 22100 | 14734/2 | 9825/9  | 6741/18 |          |        | 11.316 | 23 |
| fl1400  | 38209  | 25471 | 16981   | 11322/1 | 7551/18 |          |        | 8.087  | 16 |
| u1432   | 39979  | 26652 | 17769   | 11847   | 7899    | 5270/3   |        | 5.516  | 10 |
| fl1577  | 48487  | 32322 | 21549   | 14367/2 | 9579/10 |          |        | 9.110  | 18 |
| d1655   | 53406  | 35605 | 23737   | 15825   | 10551/2 | 7036/19  |        | 6.375  | 13 |
| vm1748  | 59577  | 39717 | 26476   | 17651   | 11768   | 7848/6   |        | 6.732  | 13 |
| u1817   | 64376  | 42917 | 28612   | 19075   | 12717   | 8479     | 5714/13| 4.666  | 9  |
| rl1889  | 69580  | 46387 | 30924/1 | 20617/3 | 13745/6 | 13745/17 |        | 10.914 | 21 |
| u2152   | 90309  | 60207 | 40139   | 26760   | 17841   | 11896/2  | 7948/9 | 5.528  | 11 |
| pr2392  | 111580 | 74387 | 49592   | 33062   | 22042   | 14695    | 9798/4 | 6.143  | 12 |

A graph with $O(n\log_2(n))$ edges is commonly considered to be sparse. The experimental results in Tables 2, 3 and 4 show us the preserved graphs with $O(n\log_2(n))$ edges contain the *OHC*. The comparisons between the number $N_k$ of edges in the preserved graph containing the *OHC* and $[n\log_2(n)]$ are shown in Table 5 where *k* is the computation cycle. We choose the $N_k$s from Tables 2, 3 and 4 where the number of the lost *OHC* edges is at most 1 in the corresponding preserved graphs. For the preserved graph excluding only one *OHC* edge, we can find the optimal path $OP^n$ in the preserved graph and it is equal to the *OHC* once its two end vertices are connected. In view of these $N_k$s and $[n\log_2(n)]$s, we find $[n\log_2(n)]$ is bigger than $N_k$ for most *TSP* instances except for a few other instances, such as brg180 and si535. Even though $N_k \geq [n\log_2(n)]$ for a few examples, the $N_k$ is close to $[n\log_2(n)]$ in values. One sees the value $\frac{N_k}{n\log_2 n}$ is very small. For example brg180 and si535, $\frac{3186}{180\log_2 180} < 2.362$ and $\frac{8180}{535\log_2 535} < 1.687$. The factors $2.362 \ll 180$ and $1.687 \ll 535$. For these special examples, the $[n\log_2(n)]$ is less than two or three times of $N_k$. It means we can compute a sparse graph with $cn = [n\log 2(n)]$ edges which includes the *OHC* for most *TSP* instances. Otherwise, we multiply $n\log_2(n)$ by a small number, such as 1.5 or 2.5, and keep such number of edges with big frequency in the preserved graph for *TSP*. Finally, the algorithm outputs a sparse graph with $O(n\log_2(n))$ edges for *TSP*.

**Table 5.** The comparisons between the number $N_k$ of edges containing the *OHC* and $n\log_2(n)$ where $k$ is the computation cycle.

| TSP | gr17 | gr21 | gr24 | fri26 | bayg29 | dantzig42 | gr48 | hk48 | brazil58 |
|---|---|---|---|---|---|---|---|---|---|
| $k$ | 4 | 5 | 5 | 5 | 4 | 5 | 5 | 5 | 6 |
| $N_k$ | 30 | 32 | 54 | 45 | 81 | 115 | 149 | 150 | 146 |
| $[n\log_2 n]$ | 69 | 92 | 110 | 122 | 141 | 226 | 268 | 268 | 340 |
| TSP | att48 | eil51 | st70 | eil76 | pr76 | berlin52 | gr96 | rat99 | kroA100 |
| $k$ | 5 | 7 | 5 | 6 | 5 | 6 | 7 | 7 | 7 |
| $N_k$ | 152 | 90 | 320 | 255 | 379 | 123 | 277 | 291 | 294 |
| $[n\log_2 n]$ | 268 | 289 | 429 | 475 | 475 | 296 | 632 | 656 | 664 |
| TSP | rd100 | eil101 | lin105 | kroB100 | kroC100 | kroD100 | pr107 | pr124 | bier127 |
| $k$ | 6 | 7 | 6 | 7 | 7 | 7 | 6 | 7 | 8 |
| $N_k$ | 440 | 300 | 485 | 294 | 295 | 296 | 500 | 452 | 320 |
| $[n\log_2 n]$ | 664 | 672 | 704 | 664 | 664 | 664 | 721 | 862 | 887 |
| TSP | ch130 | pr136 | gr137 | pr144 | kroA150 | kroB150 | ch150 | pr152 | u159 |
| $k$ | 7 | 8 | 7 | 7 | 8 | 7 | 8 | 6 | 7 |
| $N_k$ | 496 | 367 | 550 | 606 | 442 | 659 | 440 | 1010 | 740 |
| $[n\log_2 n]$ | 912 | 963 | 972 | 1032 | 1084 | 1084 | 1084 | 1101 | 1163 |
| TSP | si175 | brg180 | rat195 | d198 | kroA200 | kroB200 | gr202 | ts225 | tsp225 |
| $k$ | 6 | 4 | 9 | 8 | 8 | 8 | 8 | 8 | 9 |
| $N_k$ | 1300 | 3186 | 504 | 766 | 783 | 782 | 802 | 988 | 666 |
| $[n\log_2 n]$ | 1304 | 1349 | 1483 | 1511 | 1529 | 1529 | 1547 | 1758 | 1758 |
| TSP | pr226 | gr229 | pr264 | a280 | pr299 | lin318 | rd400 | fl417 | gr431 |
| $k$ | 7 | 8 | 8 | 9 | 8 | 8 | 8 | 9 | 10 |
| $N_k$ | 1487 | 1023 | 1360 | 1021 | 1745 | 1972 | 3119 | 2260 | 1615 |
| $[n\log_2 n]$ | 1767 | 1795 | 2124 | 2276 | 2459 | 2643 | 3458 | 3630 | 3772 |
| TSP | pr439 | d493 | si535 | att532 | pcb442 | pa561 | u574 | rat575 | p654 |
| $k$ | 9 | 10 | 7 | 10 | 9 | 9 | 10 | 10 | 9 |
| $N_k$ | 2506 | 2111 | 8180 | 2542 | 2541 | 4084 | 2855 | 2867 | 5556 |
| $[n\log_2 n]$ | 3854 | 4410 | 4848 | 4817 | 3884 | 5123 | 5260 | 5271 | 6117 |
| TSP | d657 | gr666 | u724 | pr1002 | u1060 | vm1084 | rat783 | gr120 | ail535 |
| $k$ | 10 | 10 | 11 | 11 | 11 | 10 | 11 | 7 | 9 |
| $N_k$ | 3743 | 3848 | 3031 | 5800 | 6491 | 10180 | 3547 | 419 | 3718 |
| $[n\log_2 n]$ | 6149 | 6247 | 6878 | 9989 | 10653 | 10929 | 7527 | 829 | 4849 |
| TSP | rl1304 | rl1323 | fl1400 | fl1577 | pcb1173 | d1291 | u1432 | d1655 | u1817 |
| $k$ | 9 | 9 | 11 | 10 | 11 | 11 | 12 | 11 | 13 |
| $N_k$ | 22100 | 22747 | 11322 | 21549 | 7949 | 9627 | 7899 | 15825 | 8479 |
| $[n\log_2 n]$ | 13471 | 13719 | 14632 | 16752 | 11960 | 13342 | 15013 | 17696 | 19673 |
| TSP | vm1748 | | rl1889 | | u2152 | | pr2392 | | |
| $k$ | 12 | | 10 | | 12 | | 13 | | |
| $N_k$ | 11768 | | 30924 | | 17841 | | 14695 | | |
| $[n\log_2 n]$ | 19269 | | 20559 | | 23826 | | 26848 | | |

It mentions that some *TSP* examples contain many equal-weight edges. If there are no improvements for the equal-weight edges, our iterative algorithm cannot prove to compute a sparse graph for these special *TSP* examples. Given a quadrilateral *ABCD*, it has more than 6 $OP^4$s and the corresponding frequency quadrilaterals *ABCD* are not unique. In this case, it is difficult to compute the right frequency quadrilateral for *ABCD* because the six $OP^4$s are hard to choose. In our experiments, we select the $OP^4$s in *ABCD* according to the lexicographic orders of vertices. For example in quadrilateral *ABCD*, if the two paths $OP^4$s (*A, B, C, D*) and (*A, C, B, D*) for two endpoints *A* and *D* have the equal distance, we choose the path (*A, B, C, D*) as the $OP^4$ and neglect the other path (*A, C, B, D*). The method to choose the $OP^4$s guarantees us to compute the 6 $OP^4$s for the quadrilateral *ABCD*. On the other hand, some $OP^4$s may not be right to compute the frequency of edges, especially for some *OHC* edges in $K_n$. When many wrong $OP^4$s are used to compute the frequency quadrilaterals, the frequency of edges computed with these frequency quadrilaterals do not conform to the binomial distribution model [18], which leads to the smaller frequency of some *OHC* edges in $K_n$. Thus, we may eliminate these *OHC* edges according to their frequency in the computation process.

Through computation, we found some examples, such as lin105, pr124, pr152, si175, brg180, d198, pr264, si535, pr1002, d1291, rl1304, rl1323 and rl1889, have many equal-weight edges. Moreover, they have many equal $OP^4$s (*A, B, C, D*) and (*A, C, B, D*) in the quadrilaterals *ABCD*. For example if *A* = 0 and *D* = 1 for rl1304, there are 101 pairs such $OP^4$s (0, *B, C*, 1) and (0, *C, B*, 1). For the other two endpoints, they also have many pairs of such equal $OP^4$s. In this case, it is difficult to choose the right $OP^4$s to compute the frequency of the *OHC* edges. Many wrong $OP^4$s will be chosen to compute the frequency quadrilaterals and the smaller frequency of the *OHC* edges are computed with these frequency quadrilaterals. Thus, these *OHC* edges will be eliminated according to their small average frequency at a small computation cycle.

In paper [18], the authors suggested to adding random small distances to the distances of equal-weight edges so that the special *TSP* will become the general *TSP*. Because the random distances are much smaller than the distances of edges, they generally do not change the *OHC* edges in the $K_n$. After finding the *OHC* edges based on the frequency graphs, we can compute the distance of the *OHC* with the actual distances of the edges in the $K_n$. In Tables 2 and 3, we use the method for the three special *TSP* si175, brg180 and si535. Before the execution of the iterative algorithm, we add random small distances $rd \in [0, 1]$ to the distances for all of edges. Then we run our iterative algorithm based on the new distances of edges. Although this method cannot guarantee to change the worst case into the best case every time, it usually changes the special *TSP* into a general *TSP* with a high probability in many experiments. The experiments in paper [18] have proven the function of the random small distances so that our probability model can works for the special *TSP*.

According to the experimental results, we cannot execute the iterative algorithms $\left\lfloor log_{\frac{2}{3}}\left(\frac{2}{n-1}\right) \right\rfloor$ times to compute a sparse graph with *n* edges for *TSP*. There are two reasons. One reason is that there exist the special *TSP* instances with many equal-weight edges. The average frequency of the *OHC* edges does not conform to the probability model unless they are transformed into the general *TSP*. The second reason is when the number *N* of quadrilaterals containing every edge becomes very small, our probability model does not work efficiently. Therefore, we should stop the iterative algorithm before the maximum computation cycle $\left\lfloor log_{\frac{2}{3}}\left(\frac{2}{n-1}\right) \right\rfloor$. For general *TSP*, the average frequency of the *OHC* edges will be bigger than the average frequency of all of edges in the preserved graphs in the previous computation cycles. When we abandon the $\frac{1}{3}|E_{k-1}|$ edges at each computation cycle, we expect the probability of all edges meet the probability condition $P(f \leq 3) > 1/3$ at each computation cycle. Thus, we take the $P(f \leq 3) < 1/3$ as a restrictive condition to terminate the com-putation process. In application, we will execute the iterative algorithm until the frequency of edges in the preserved graph meets the probability $P(f \leq 3) \leq 1/3$. The last computation cycle where $P(f \leq 3) \geq 1/3$ to compute the residual graph is taken as the stop cycle $k_s$. At the stop computation cycle, we will obtain a sparse graph with $\left[\left(\frac{2}{3}\right)^{k_s} \binom{n}{2}\right]$ edges for the *TSP*.

To verify the feasibility of the terminal condition, we did the experiments for the *TSP* instances in Tables 2, 3 and 4 to analyze the corresponding sparse graphs at the $k_s^{th}$ computation cycle. In these experiments, we add random small distances $rd \in [0, 1]$ to all of edges for every example. For the special *TSP* instances, the equal-weight edges will be greatly reduced.

Meanwhile, the results will be a little different from those in Tables 2, 3 and 4 due to the effect of the random small distances. In these experiments, we recorded the stop computation cycle $k_s$ and the number of edges $|E_{k_s}|$ in the corresponding preserved graphs. The number of edges with average frequency less than 3 in $E_{k_s-1}$ is computed and noted as $|E_{f<3}|$. The values $\left[\frac{1}{3}|E_{k_s-1}|\right]$ is computed for comparison. The parameters $c$ and average degree $d$ of the sparse graphs are computed according to $|E_{k_s}|$. The number of the lost *OHC* edges is also recorded as $l_{ohc}$. If $l_{ohc} = 1$, we assume this sparse graph owns the *OHC*. The results for these examples are shown in Table 6.

At the $k_s^{th}$ cycle, the preserved graphs include the *OHC* for almost all of the *TSP* instances. It is obvious that the preserved graphs include the *OHC* for these instances before the stop computation cycle $k_s$. According to the results in Tables 2, 3 and 4, a few more iterations can be executed after the $k_s^{th}$ computation cycle and another even sparser graph with the *OHC* will be generated for most of these instances. For example rat783, the preserved graph still contains the *OHC* at the $11^{th}$ computation cycle whereas the stop computation cycle is 8.

Moreover, we see that the stop computation cycle $k_s$ increases according to the scale of *TSP* $n$. In general case, the larger the scale $n$ of *TSP* is, the bigger the stop computation cycle is. It means we can iterate the algorithm several more times for larger scale of *TSP* and find a sparse graph for them. Since the number of edges in the preserved graphs decreases in proportion to 2/3, we will compute a further sparser graph if the iterative algorithm is executed a few more times.

In addition, we see that some *TSP* instances have the same stop cycle $k_s$ when their scales do not have much difference or in one interval. For the instances pr144 and gr229, their $k_s$s is equal to 6 but their scales are 144 and 229, respectively. At the same stop cycle $k_s$, the preserved graph of the bigger *TSP* will have the bigger parameters $c$ and $d$. It seems for the bigger *TSP* we cannot compute the residual graph as sparse as that for a smaller *TSP*. For different scale of *TSP* instances, it mentions that the difference between $|E_{f<3}|$ and $\frac{1}{3}|E_{k-1}|$ becomes bigger and bigger according to $n$ at the same stop cycle $k_s$. It means that the number of edges with the average frequency less than 3 increases according to $n$ at the same stop cycle $k_s$. Originally, we should throw away more edges rather than $\frac{1}{3}|E_{k-1}|$ edges for larger scale of *TSP* at the same computation cycle. To eliminate more number of edges with the average frequency less than 3 in $E_{k-1}$, the algorithm has to run more computation cycles for the larger

*TSP* instances. Thus, an even sparser graph than that computed at the stop computation cycle will be generated. For example pr144 and gr229 in Table 2 and 3, the algorithm iterates two more times for gr229 whereas it runs one more time for pr144 after the stop computation cycle. At last, in the two final preserved graphs of pr144 and gr229, the parameters *c* and *d* are nearly equal. When the scale of *TSP* is beyond an appointed value, the stop computation cycle $k_s$ will increase accordingly. For example pr264, the stop computation cycle $k_s$ becomes 7.

We are interested in the other problem: does the preserved graphs include the *OHC* once $\frac{1}{3}|E_{k-1}|$ is bigger than $|E_{f<3}|$ in the computation process? We discuss the question because we compute the frequency of edges with the incomplete frequency quadrilaterals at the late stage. Although the average frequency of the *OHC* edges will be less than 3, we hope they are bigger than that of the 1/3 edges to be eliminated. If the preserved graph includes the *OHC*, we could execute the iterative algorithm at least one more time after the stop computation cycle to generate the sparser graph for *TSP*. Here we use a ratio computed as $\rho = \frac{\frac{1}{3}|E_{k_s}| - |E_{f<3}|}{\frac{1}{3}|E_{k_s}|}$ to show the difference between $\frac{1}{3}|E_{k_s}|$ and $|E_{f<3}|$. We will execute the iterative algorithm once when we meet $\frac{1}{3}|E_{k_s}| > |E_{f<3}|$ for the first time. Meanwhile, the computation cycle and $\rho$ are computed. The number of the lost *OHC* edges are also recorded in the corresponding residual graph. The results are given in Table 7. Comparing with the results in Table 6, we see most of these preserved graphs still include the *OHC* at the $(k_s + 1)^{th}$ computation cycle. It means we can execute the iterative algorithm one more time after the stop computation cycle $k_s$ even though $\frac{1}{3}|E_{k_s}| > |E_{f<3}|$. Meanwhile, a further sparser graph is generated for *TSP*. We also find for some special *TSP* instances, such as si175, brg180 and si535, the value $\rho$ is near to 1 at the $(k_s+1)^{th}$ computation cycle. It says the $\frac{1}{3}|E_{k_s}|$ is much bigger than $|E_{f<3}|$. Therefore, many edges with the average frequency bigger than 3 will be eliminated at this computation cycle. The residual graphs will lose several or quite a few *OHC* edges. For the special *TSP* instances with many equal-weight edges, we suggest the sparse graphs computed at the stop computation cycle $k_s$ are more suitable for *TSP*.

We have two counter examples brg180 and si535. The preserved graphs lose quite a few *OHC* edges at the stop cycle. The main reason is that the brg180 and si535 have many equal-weight edges. Given a set of *k* vertex, there are more than one *OHC*s and $\binom{k}{2}$ optimal *k*-vertex paths. For brg180, we use the previous 10 vertices and the corresponding distances to construct a small *TSP*. It finds the small *TSP* has 128 *OHC*s. Even though we add random small distances to brg180 and si535, the random distances cannot ensure the edges in the *OHC* to have the big frequency in the corresponding frequency quadrilaterals. After all, the random small distances only play the average effect to compute the frequency for all of edges. Moreover, we cannot guarantee the *OHC* edges to have a big frequency in one experiment. In our experiments, some *OHC* edges may have a small frequency in their frequency quadrilaterals. Therefore, these *OHC* edges will be abandoned before the stop computation cycle. If we did several or many experiments, the better result would be computed.

**Table 6.** The computational results at the stop computation cycle

| TSP | $k_s$ | $\frac{1}{3}|E_{k_s}|/|E_{f<3}|$ | $l_{ohc}$ | c/d | TSP | $k_s$ | $\frac{1}{3}|E_{k_s}|/|E_{f<3}|$ | $l_{ohc}$ | c/d |
|---|---|---|---|---|---|---|---|---|---|
| gr17 | 3 | 21/23 | 0 | 2.529/5 | gr202 | 6 | 891/1852 | 0 | 8.832/18 |
| gr21 | 3 | 31/37 | 0 | 3.048/6 | ts225 | 6 | 1106/2459 | 0 | 9.840/20 |
| gr24 | 3 | 41/47 | 0 | 3.458/7 | tsp225 | 6 | 1106/2275 | 0 | 9.840/20 |
| fri26 | 3 | 48/94 | 0 | 3.731/7 | pr226 | 6 | 1116/1854 | 1 | 9.845/20 |
| bayg29 | 3 | 60/119 | 0 | 4.172/8 | gr229 | 6 | 1146/2373 | 0 | 10.013/20 |
| dantzig42 | 3 | 128/251 | 0 | 6.095/12 | pr264 | 7 | 1016/2194 | 1 | 7.701/15 |
| gr48 | 4 | 111/221 | 0 | 4.625/9 | a280 | 7 | 1143/2410 | 0 | 8.168/16 |
| hk48 | 4 | 112/228 | 0 | 4.667/9 | pr299 | 7 | 1304/2732 | 0 | 8.716/17 |
| att48 | 4 | 112/217 | 0 | 4.667/9 | lin318 | 7 | 1475/3060 | 0 | 9.296/19 |
| eil51 | 5 | 85/85 | 0 | 3.373/7 | rd400 | 7 | 2336/4841 | 0 | 11.683/23 |
| berlin52 | 4 | 131/265 | 0 | 5.077/10 | fl417 | 7 | 2538/5051 | 0 | 12.177/24 |
| brazil58 | 4 | 163/326 | 0 | 5.603/11 | gr431 | 7 | 2712/5645 | 0 | 12.587/25 |
| st70 | 4 | 239/494 | 0 | 6.843/14 | pr439 | 7 | 2814/5818 | 0 | 12.822/26 |
| eil76 | 4 | 282/567 | 0 | 7.434/15 | pcb442 | 7 | 2852/5891 | 0 | 12.907/26 |
| pr76 | 4 | 282/585 | 0 | 7.421/15 | d493 | 8 | 2366/4896 | 0 | 9.602/19 |
| gr96 | 5 | 301/630 | 0 | 6.302/13 | att532 | 8 | 2756/5696 | 0 | 10.359/21 |
| rat99 | 5 | 320/660 | 0 | 6.475/13 | ail535 | 8 | 2787/5921 | 0 | 10.422/21 |
| kroA100 | 5 | 326/669 | 0 | 6.530/13 | si535 | 8 | 2727/6059 | 8 | 9.907/20 |
| kroB100 | 5 | 327/686 | 0 | 6.560/13 | pa561 | 8 | 3063/6396 | 0 | 10.906/22 |
| kroC100 | 5 | 326/672 | 0 | 6.530/13 | u574 | 8 | 3208/6711 | 0 | 11.179/22 |
| kroD100 | 5 | 326/655 | 0 | 6.530/13 | rat575 | 8 | 3220/6611 | 0 | 11.200/22 |
| rd100 | 5 | 326/655 | 0 | 6.540/13 | p654 | 8 | 4165/8579 | 0 | 12.737/25 |
| eil101 | 5 | 333/662 | 0 | 6.604/13 | d657 | 8 | 4205/8805 | 0 | 12.802/26 |
| lin105 | 5 | 359/737 | 0 | 6.876/14 | gr666 | 8 | 4321/9118 | 0 | 12.977/26 |
| pr107 | 5 | 373/775 | 0 | 6.972/14 | u724 | 8 | 5106/10625 | 0 | 14.108/28 |
| gr120 | 5 | 470/934 | 0 | 7.825/16 | rat783 | 8 | 5973/12297 | 0 | 15.258/31 |
| pr124 | 5 | 502/1060 | 1 | 8.105/16 | pr1002 | 9 | 6523/13653 | 0 | 13.380/27 |
| bier127 | 5 | 527/1099 | 0 | 8.307/17 | u1060 | 9 | 7301/15243 | 0 | 13.776/28 |
| ch130 | 5 | 552/1143 | 0 | 8.508/17 | vm1084 | 9 | 7634/15733 | 0 | 14.086/28 |
| pr136 | 5 | 605/1206 | 0 | 8.904/18 | pcb1173 | 9 | 8940/18494 | 0 | 15.245/30 |
| gr137 | 5 | 613/1260 | 0 | 8.964/18 | d1291 | 9 | 10830/22808 | 0 | 16.777/33 |
| pr144 | 6 | 452/990 | 0 | 6.285/13 | rl1304 | 9 | 11050/24286 | 0 | 16.974/34 |
| ch150 | 6 | 491/1014 | 0 | 6.553/13 | rl1323 | 9 | 11374/24631 | 1 | 17.193/34 |
| kroA150 | 6 | 491/1009 | 0 | 6.560/13 | fl1400 | 9 | 12736/26223 | 0 | 18.193/36 |
| kroB150 | 6 | 491/1004 | 0 | 6.553/13 | u1432 | 9 | 13326/27704 | 0 | 18.612/37 |
| pr152 | 6 | 503/1072 | 0 | 6.632/13 | fl1577 | 9 | 16162/34902 | 0 | 20.496/41 |
| u159 | 6 | 552/1147 | 0 | 6.950/14 | d1655 | 10 | 11868/25141 | 0 | 14.343/29 |
| si175 | 6 | 665/1447 | 0 | 7.434/15 | vm1748 | 10 | 13239/27647 | 0 | 14.841/30 |
| brg180 | 6 | 707/1513 | 19 | 7.133/14 | u1817 | 10 | 14305/29859 | 0 | 15.746/31 |
| rat195 | 6 | 831/1639 | 0 | 8.528/17 | rl1889 | 10 | 15462/34122 | 1 | 16.370/33 |
| d198 | 6 | 856/1759 | 0 | 8.657/17 | | | | | |
| kroA200 | 6 | 874/1798 | 0 | 8.745/17 | u2152 | 10 | 20069/42176 | 0 | 18.652/37 |
| kroB200 | 6 | 874/1795 | 0 | 8.745/17 | pr2392 | 10 | 24795/52301 | 0 | 20.732/41 |

**Table 7.** The computational results when $\frac{1}{3}|E_{k_s}| > |E_{f<3}|$ for the first time

| TSP | k | $\frac{1}{3}|E|/|E_{f<3}|$ | $l_{ohc}$ | d | TSP | k | $\frac{1}{3}|E|/|E_{f<3}|$ | $l_{ohc}$ | d |
|---|---|---|---|---|---|---|---|---|---|
| gr17 | 4 | 14/14 | 1 | 4 | gr202 | 7 | 594/399 | 0 | 12 |
| gr21 | 4 | 21/19 | 0 | 4 | ts225 | 7 | 738/149 | 0 | 13 |
| gr24 | 4 | 27/24 | 1 | 5 | tsp225 | 7 | 738/582 | 0 | 13 |
| fri26 | 4 | 32/28 | 0 | 5 | pr226 | 7 | 741/428 | 3 | 13 |
| bayg29 | 4 | 40/35 | 1 | 5 | gr229 | 7 | 764/535 | 0 | 13 |
| dantzig42 | 5 | 57/45 | 0 | 5 | pr264 | 8 | 677/304 | 1 | 10 |
| gr48 | 5 | 74/60 | 0 | 6 | a280 | 8 | 762/454 | 0 | 11 |
| hk48 | 5 | 74/71 | 0 | 6 | pr299 | 8 | 869/592 | 1 | 12 |
| att48 | 5 | 74/65 | 0 | 6 | lin318 | 8 | 985/556 | 0 | 12 |
| eil51 | 6 | 57/ 43 | 1 | 5 | rd400 | 8 | 1557/1169 | 0 | 16 |
| berlin52 | 5 | 88/64 | 0 | 7 | fl417 | 8 | 1692/1041 | 0 | 16 |
| brazil58 | 5 | 108/59 | 0 | 7 | gr431 | 8 | 1808/1325 | 0 | 17 |
| st70 | 5 | 159/134 | 0 | 9 | pr439 | 8 | 1876/1152 | 0 | 17 |
| eil76 | 5 | 188/186 | 0 | 10 | pcb442 | 8 | 1901/1488 | 0 | 17 |
| pr76 | 5 | 188/157 | 1 | 10 | d493 | 9 | 1578/1080 | 0 | 13 |
| gr96 | 6 | 201/133 | 0 | 8 | att532 | 9 | 1837/1250 | 0 | 14 |
| rat99 | 6 | 213/182 | 0 | 9 | ail535 | 9 | 1858/826 | 0 | 14 |
| kroA100 | 6 | 217/159 | 0 | 9 | si535 | 9 | 1766/57 | 13 | 13 |
| kroB100 | 6 | 218/164 | 0 | 9 | pa561 | 9 | 2039/603 | 0 | 15 |
| kroC100 | 6 | 217/159 | 1 | 9 | u574 | 9 | 2139/1429 | 0 | 15 |
| kroD100 | 6 | 217/163 | 0 | 9 | rat575 | 9 | 2146/1706 | 0 | 15 |
| rd100 | 6 | 218/168 | 0 | 9 | p654 | 9 | 2776/1612 | 0 | 17 |
| eil101 | 6 | 222/185 | 0 | 9 | d657 | 9 | 2803/1926 | 0 | 17 |
| lin105 | 6 | 240/166 | 1 | 9 | gr666 | 9 | 2881/1718 | 0 | 17 |
| pr107 | 6 | 248/188 | 0 | 9 | u724 | 9 | 3404/2661 | 0 | 19 |
| gr120 | 6 | 314/206 | 0 | 10 | rat783 | 9 | 3982/3348 | 0 | 20 |
| pr124 | 6 | 335/222 | 1 | 11 | pr1002 | 10 | 4349/2853 | 0 | 17 |
| bier127 | 6 | 351/245 | 0 | 11 | u1060 | 10 | 4867/3259 | 0 | 18 |
| ch130 | 7 | 246/183 | 1 | 8 | vm1084 | 10 | 5089/3895 | 0 | 19 |
| pr136 | 7 | 269/205 | 0 | 8 | pcb1173 | 10 | 5960/4518 | 1 | 20 |
| gr137 | 7 | 273/190 | 0 | 8 | d1291 | 10 | 7219/4265 | 1 | 22 |
| pr144 | 7 | 301/92 | 0 | 8 | rl1304 | 10 | 7366/3843 | 2 | 23 |
| ch150 | 7 | 327/211 | 0 | 9 | rl1323 | 10 | 7582/4301 | 5 | 23 |
| kroA150 | 7 | 328/230 | 0 | 9 | fl1400 | 10 | 8490/4957 | 0 | 24 |
| kroB150 | 7 | 327/238 | 0 | 9 | u1432 | 10 | 8884/7391 | 0 | 25 |
| pr152 | 7 | 336/129 | 3 | 9 | fl1577 | 10 | 10774/4315 | 0 | 27 |
| u159 | 7 | 368/249 | 0 | 9 | d1655 | 11 | 7912/4826 | 0 | 19 |
| si175 | 7 | 434/33 | 3 | 10 | vm1748 | 11 | 8825/6184 | 0 | 20 |
| brg180 | 7 | 431/11 | 21 | 10 | u1817 | 11 | 9537/69323 | 0 | 21 |
| rat195 | 7 | 554/477 | 0 | 11 | rl1889 | 11 | 10308/4632 | 3 | 22 |
| d198 | 7 | 571/363 | 1 | 12 | | | | | |
| kroA200 | 7 | 583/424 | 0 | 12 | u2152 | 11 | 13379/10368 | 0 | 25 |
| kroB200 | 7 | 583/413 | 0 | 12 | pr2392 | 11 | 16530/11978 | 0 | 28 |

## 5 Conclusions

We design a heuristic algorithm based on the frequency quadrilaterals to compute the sparse graph for *TSP*. When the frequency of an edge *e* is computed with *N* frequency quadrilaterals containing *e*, the frequency of the *OHC* edges will be bigger than the average frequency 3*N* of all of edges when *N* is big enough. The probability model shows it is likely $\left\lfloor \frac{2}{3}|E| \right\rfloor$ edges whose frequency is above the average frequency 3. Thus, we can eliminate $\left\lfloor \frac{1}{3}|E| \right\rfloor$ edges with low frequency so as to compute a preserved graph for *TSP*. We iterate the elimination process until a sparse graph is obtained for *TSP* as *N* is big enough in the preserved graphs. In an ideal case, a sparse graph with *cn* edges is computed at the $\left\lfloor log_{\frac{2}{3}}\left(\frac{2c}{n-1}\right) \right\rfloor^{th}$ computation cycles where $c = log_2 n$. We tested the algorithm with tens of various *TSP* instances. The experimental results showed that our probability model works well for general *TSP* instances. The sparse graphs with $O(n\log_2(n))$ edges are computed for these instances. It says the average frequency of the *OHC* edges is bigger than that of the 1/3 edges to be eliminated not only in the $K_n$, but also in the preserved graphs that the algorithm computes. Thus, the *OHC* edges are always preserved in the computation process until the stop computation cycle is arrived.

In the near future, the properties of the residual graphs will be analyzed. We expect the sparse graphs have the good properties, such as bounded degree, genus, tree-width and planarity, etc. so that we can design the polynomial-time algorithms or polynomial-time approximation algorithms for *TSP* based on the sparse graphs. In addition, the other terminal conditions will be explored to compute the sparse graphs with good properties.

## 6 Acknowledgement

The authors acknowledge the anonymous referees for their suggestions to improve the paper. We acknowledge W. Cook, H. Mittelmann who created the Concorde and G. Reinelt, et al. who provided the *TSP* data to *TSPLIB*. The authors acknowledge the funds supported by NSFC (No.51205129) and the Fundamental Research Funds for the Central Universities (No.2015ZD10). We also thank the support of Beijing Key Laboratory of Energy Safety and Clean Utilization.